\documentclass[twocolumn]{aastex631}

\usepackage{amsmath}

\begin{document}

\title{Chromaticity Effects on the Outcomes of Spheroid-based Scored Events}

\author[0000-0002-0786-7307]{Lisa McBride}
\affiliation{Trottier Space Institute \\
McGill University \\
3550 Rue University, Montréal, QC H3A 2A7}

\author{Michael Pagano}
\affiliation{Trottier Space Institute \\
McGill University \\
3550 Rue University, Montréal, QC H3A 2A7}



\begin{abstract}

The immense popularity of spheroid-based scored events (colloquially ``football games'') motivates the desire to better understand the underlying mechanisms affecting their outcomes. By construction of these events, participants must distinguish the spheroidal ball from not only the background, but also their team and enemy players, which are marked by self-assigned linear combinations of specific frequencies of electromagnetic magnetic radiation, known as uniform color. We investigate chromatic effects on the outcome of such events. We do this by finding the correlation between the color contrast and the success of several key spheroidal ball match tactics. We perform this analysis for the 2020 NFL regular season, focusing on moves in which uniform colors may be a factor in performance.  We conduct a primary analysis using each team's cumulative results over the season, but in doing so neglect non-uniformity in the chosen uniform color per individual. We then conduct a secondary analysis of the performance per game of a single team, the Seattle Seahawks, which exhibited large uniform color variability for the 2020 NFL regular season.  In this work, tackles and completions are considered. The Pearson correlation coefficient is then calculated for both tactics. We find little evidence of chromaticity effects, with correlation values of $r_t=-0.0885\pm 0.1819$ and $r_c-0.0292\pm0.1825 $, respectively, for the primary analysis. 
\end{abstract}



\section{Introduction} \label{sec:intro}
Competitive ranked group-based events are a common modality for both fun and profit in much of the world, having been observed at nearly all longitudes and latitudes. In most cases these events involve a binary form of play, where the participating contestants are split into two discrete groups, or teams.  These teams are then paired with an opposing team in order to determine the relative skill of each team (or rather produce one particular realization of the underlying skill distributions).  These tests are often referred to in the literature as games.

One requirement of such events is the need for unique, identifying markers between both competing parties.  This is necessary both to provide narrative accessibility for observers, but also as critical visual information for \textit{in situ} decision making by the competitors.  The dominant strategy for such identification is the use of unique linear combinations of frequencies of the electromagnetic spectrum, called team colors.

The outcome of each game depends heavily on the body response of the contestants.  Chief among body responses is hand-eye coordination.  This coordination is critical to achieving a good result. Due to the chaotic nature of a game, many decisions must be made in real time, requiring the last minute processing of visual data \citep{ballard1992hand}. In particular, the dynamic response of the human eye in the time-domain affects the body's coordination, which affects the body's ability to track moving targets, a critical skill in matches \citep{gonzalez2016effects, mrotek2007target}.
 
It is known that the frequency response of the human eye is non-uniform \cite{goldsmith1990optimization}. Additionally, this frequency response can affect not only the attention, but also the subsequent hand-eye coordination of the subject \citep{serences2005coordination}.  Research suggests that eye fixation is correlated with the color contrast of the input \citep{tatler2005visual, chen2017enhanced, nasanen2001effect}. 

While the need to visually track players in real time is necessary for all team events, it is especially critical for American football (referred to in this work from hereonin as football, despite the inaccuracy), where the players themselves are tracked as intently as the target object, a spheroid MacGuffin known as a football.

In football, two teams are placed upon a large Cartesian grid, taking turns to travel from one side to the other.  During a turn, the team chooses an internal member of the group to receive the ball.  This is done by passing the chosen player the spheroid via the use of projectile motion.  The player then runs as far as they can without falling over.  Once fallen (or down), the attempt is ended.  Each team gets four attempts of running without falling over before their turn is ended. Once a turn is completed, the opposing team receives the spheroid ball and their turn begins. 

However, the player running down the field is, in general, not allowed to free stream. Instead, their mean free path depends on both the player's four-momentum, $p^{\alpha}$, and the opposing team's ability to arrest that momentum with a precise application of conservation of momentum.  Because player's themselves are tracked as intently as the ball itself, it is possible that chromaticity effects between uniforms will have a disproportionatley greater impact than on other team-based events in which the ball is the primary target.

Uncertainties remain on the effect of differing color templates of team uniforms on the outcomes of scored events.  We investigate to what extent the results are potentially perturbed by such chromaticity effects.  We begin with a brief introduction to the models necessary for the analysis in Section\,\ref{sec:background}.  In Section\,\ref{sec:methodology} we describe the various data sets used in this work, along with a description of our analysis techniques.  Results are presented in Section\,\ref{sec:results}.  Finally, we make wild statements about following up on loose ends despite knowing that we have no intention of actually doing so in Section\,\ref{sec:conclusion}. It also has a summary.

\section{Background} \label{sec:background}
Th investigation herein is performed on data from games sponsored by the National Football League (NFL), and are confined to the 2020 regular season. Thus, our teams are restricted to official NFL teams such that for a team, i, $\{i | i \in \mathrm{NFL} \}$.   All colors described are given in RGB values.
\subsection{Team Colors}
All teams in the target group, NFL, are assigned a unique set of identifying colors. These colors can be divided broadly into two distinct categories depending on the magnitude of the spacetime vector, $\vec{s}_i$,\footnote{Note that the subscript here does not indicate a Einstein vector.} where the origin of the coordinate system, $i$, is defined to be the spacetime coordinates of the team's affiliated expensive lawn fortress, known as a stadium.  If the spacetime invariant, $\Delta s^2_i = 0$, then the match is designated as home, and similarly is designated as away if $\Delta s^2_i > 0$.  A list of all the colors used in this analysis are contained in Table\,\ref{table:colors}.
\subsubsection{Home Colors}
All teams have a selection of colors that are uniquely assigned.  However, one color is present with a larger surface area than others. This main color is the primary for each uniform.  We assign this as our home color, and neglect additional chromaticity effects of the secondary colors as they are likely to be subdominant.

\begin{deluxetable}{lcc}
\tabletypesize{\footnotesize}
\tablecolumns{3}
\tablewidth{0pt}
\tablecaption{Team Colors \label{table:colors}}
\tablehead{
\colhead{Name} \vspace{-0.2cm} & \colhead{Home Color} & \colhead{Away Color}  \\}
\startdata
\vspace{-0.0cm} Arizona Cardinals & (151,35,63) & (1,1,1)  \\
\vspace{-0.0cm} Atlanta Falcons & (167,25,48) & (1,1,1)  \\
\vspace{-0.0cm} Baltimore Ravens & (26,25,95) & (1,1,1)  \\
\vspace{-0.0cm} Buffalo Bills & (0,51,141) & (1,1,1)  \\
\vspace{-0.0cm} Carolina Panthers & (0,133,202) & (1,1,1)  \\
\vspace{-0.0cm} Chicago Bears& (11,22,42) & (1,1,1)  \\
\vspace{-0.0cm} Cincinnati Bengals& (251,79,20) & (1,1,1)  \\
\vspace{-0.0cm} Cleveland Browns& (49,29,0) & (1,1,1)  \\
\vspace{-0.0cm} Dallas Cowboys & (0,53,148) & (1,1,1)  \\
\vspace{-0.0cm} Denver Broncos  & (251,79,20) & (1,1,1)  \\
\vspace{-0.0cm} Detroit Lions  & (0,118,182) & (1,1,1)  \\
\vspace{-0.0cm} Green Bay Packers  & (24,48,40) & (1,1,1)  \\
\vspace{-0.0cm} Houston Texans& (3,32,47) & (1,1,1)  \\
\vspace{-0.0cm} Indianapolis Colts & (0,44,95) & (1,1,1)  \\
\vspace{-0.0cm} Jacksonville  Jaguars& (16,24,32) & (1,1,1)  \\
\vspace{-0.0cm} Kansas City Chiefs  & (227,24,55) & (1,1,1)  \\
\vspace{-0.0cm} LA Rams  & (0,53,148) & (1,1,1)  \\
\vspace{-0.0cm} LA Chargers  & (0,128,198) & (1,1,1)  \\
\vspace{-0.0cm} Las Vegas Raiders  & (0,0,0) & (1,1,1)  \\
\vspace{-0.0cm} Miami Dolphins  & (0,142,151) & (1,1,1)  \\
\vspace{-0.0cm} Minnesota Vikings  & (79,38,131) & (1,1,1)  \\
\vspace{-0.0cm} New England Patriots  & (0,34,68) & (1,1,1)  \\
\vspace{-0.0cm} New Orleans Saints  & (211,188,141) & (1,1,1)  \\
\vspace{-0.0cm} New York Giants  & (1,35,82) & (1,1,1)  \\
\vspace{-0.0cm} New York Jets  & (18,87,64) & (1,1,1)  \\
\vspace{-0.0cm} Philadelphia Eagles  & (0,76,84) & (1,1,1)  \\
\vspace{-0.0cm} Pittsburgh Steelers  & (255,182,18) & (1,1,1)  \\
\vspace{-0.0cm} Seattle Seahawks  & (0,34,68) & see $\S$\,\ref{sec:seahawks}  \\
\vspace{-0.0cm} San Francisco 49'ers & (170,0,0) & (1,1,1)  \\
\vspace{-0.0cm} Tampa Bay Buccaneers  & (213,10,10) & (1,1,1)  \\
\vspace{-0.0cm} Tennessee Titans & (12,35,64) & (1,1,1)  \\
\vspace{-0.0cm} Washington Commanders & (90,20,20) & (1,1,1)  \\
\enddata
\vspace{-0.8cm}
\tablecomments{Comments here}
\end{deluxetable}

\subsubsection{Away Colors}
While not an inviolable law à la conservation of energy, most teams follow the usage of white uniforms when playing away from their home stadium. However, there does exist exceptions. These exceptions are a known systematic and are expected to bias our result. However, for an initial analysis we neglect any effect due to this bias.  We revisit this assumption in Section\,\ref{sec:seahawks}

\subsubsection{Field Color}
All matches are conducted on designated surfaces, or fields, of $N=2$ dimensional rectangles composed either of natural grass, or artificial turf. The distribution of surface type per stadium is roughly bimodal with equal amplitudes \cite{grass}.

All grass and turf are expected to have spectra with peak values in the $500-550$\,nm range.  However, we expect some variability to exist between fields.  We neglect this variability for the following analysis, setting our fiducial values to be RGB$=(96,185,34)$.

\subsection{Contrast}
We model the contrast between two colors by using the WCAG guidelines for web content \citep{W3}. In this model, the contrast ratio, $C$, between two colors is given by the ratio of their relative luminances,
\begin{equation}
\label{eq:contrast_eq}
       C = \frac{L_1 + 0.05}{L_2 + 0.05}
\end{equation}
where $L_1$ ($L_2$) is the relative luminance of the lighter (darker) color. The relative luminance is defined in terms of the sRGB colorspace as,
\begin{equation}
\label{eq:contrast_eq}
L_i = 0.2126 \mathrm{R} + 0.7152 \mathrm{G} + 0.0722 \mathrm{B}
\end{equation}
where the RGB values themselves have been normalized via the prescription
\begin{equation}
\label{eq:contrast_eq}
    \Gamma = 
\begin{cases}
    \Gamma_{\mathrm{sRGB}},& \text{if } x\leq 0.03928 \\
    \left( \frac{\Gamma_{\mathrm{sRGB}} + 0.055}{1.055} \right)^{2.4},              & \text{otherwise}
\end{cases}
\end{equation}
where $\Gamma=\{\mathrm{R,G,B}\}$, $i={\{1,2\}}$, and $\mathrm{sRG}=\mathrm{RGB} / 255$.

\begin{figure*}[ht!]
\plotone{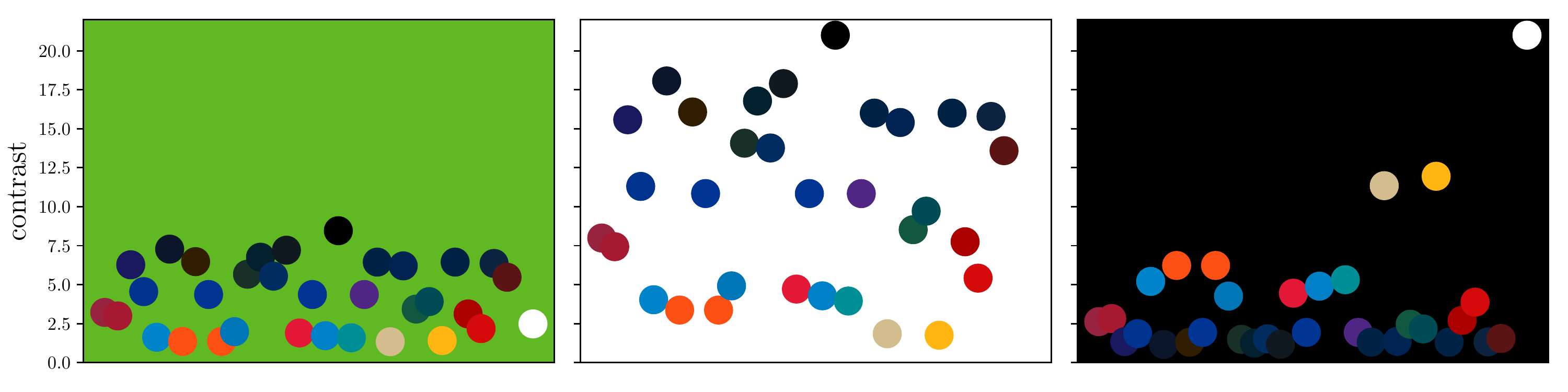}
\caption{Calculated contrasts (of a team color against a background color) of team colors (colored circles) overlaid atop the background color.  The contrast values considered in the analysis are for a field green background (left panel). Contrasts against a white (black) background are shown in the middle (right) panel for comparison.  The contrast range in the field case is smaller, as expected.
\label{fig:contrasts}}
\end{figure*}

\subsection{Tactics}
While it is possible that contrast differences can affect all forms of play, they are likely subdominant for most of the allotted time, and are complex to model.  We therefore consider two specific tactics consistently observed in spheroidal ball scored events, and examine the effects of color independently for each case. We postulate that a reduced contrast will make tackling a given player more challenging, providing an advantage to the offensive players trying escape being knocked over.  Conversely, a reduced contrast may also make completions more difficult, a detriment to the offensive team. We assume no covariance between the two categories.

\subsubsection{Tackles} \label{sec:tackles_FOOTBALL!!!!!}
The intent of the spheroidal ball scored events is to advance the spheroidal ball into the defense's territory. As such, it is the purpose of the opposing team's defense to dispossess an opponent of the ball before reaching the end goal. The main technique employed by the defense of the opposition is tackling. A tackle is said to have occurred if the defense has successfully lowered the $y$-position of any element above the foot of the ball carrier to the ground at $y=0$. A successful tackle first requires identifying the position $\textbf{x}$ of the spheroidal ball carrier, which itself requires distinguishing between the colors of the ball carrier and the field on which the scored event takes place. Thus the team colors of the ball carrier can affect the observability of the spheroidal ball with respect to the underlying field. We thus expect that ball carrier with team colors of a lower contrast with the field as defined by Equation \ref{eq:contrast_eq} will lead to a decrease in tackling. 

\subsubsection{Completions}
As stated above, objective of each scored event is to advance the spheroidal ball into the configuration space of the opposition. There are two main ways in which this can occur, by physically transporting the spheroidal ball through the defense or by throwing the spheroidal ball thought the air. In the case where the spheroidal ball is thrown, there are three possible outcomes, a completed pass, an incomplete pass, or an interception. A completion is said to have occurred when an eligible receiver of a spheroidal ball successfully catches a forward pass without having the ball reaching $y=0$. Unlike tackling from Section \ref{sec:tackles_FOOTBALL!!!!!}, a high contrast team color can potentially increase the probability of a completed pass as the target player is easier to observe. Thus, a high contrast could prove beneficial during play.

\begin{figure*}[ht!]
\plotone{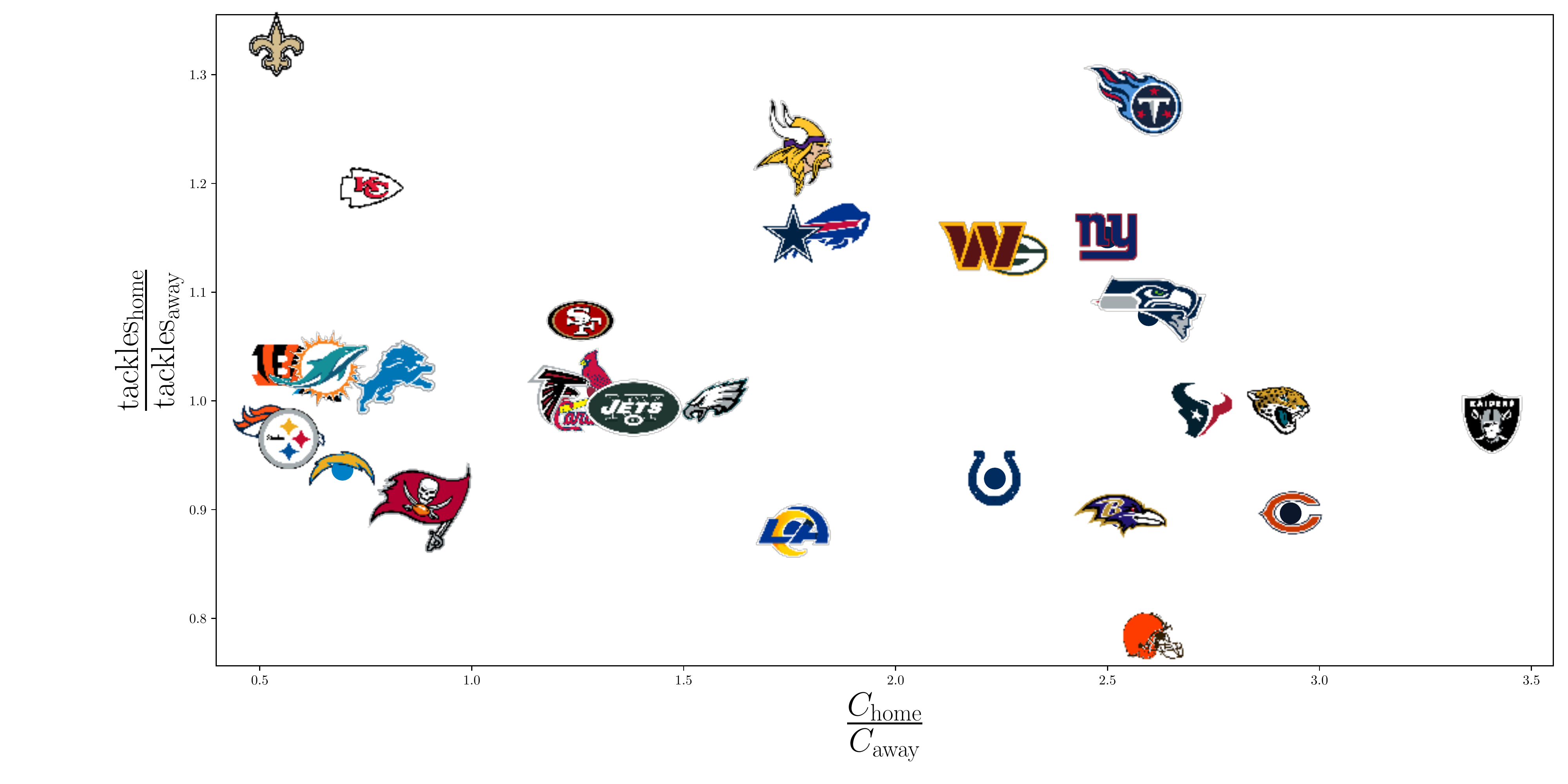}
\caption{Ratio of home to away tackles plotted as a function of the home to away color contrasts. Tackles in this context is the number of tackles performed on each team, by their opponent. Note, this is different from tackles performed by the team, which is not what is considered here.  Individual teams are plotted using their logos. 
\label{fig:tackles}}
\end{figure*} 

\section{Methodology} \label{sec:methodology}
We utilise NFL data from the 2020 season beginning on $\mathrm{JD}=2459102.50$ and ending on $\mathrm{JD}=2459217.50$.  The 2020 season, while anomalous, was chosen for our analysis due to reduced audience participation because of the global pandemic.  This reduces the known effect of home advantage (HA). 

Home advantage is a well known effect in sports \citep{nevill1999home, legaz2013home, edwards1979home} where performance is increased in a statistically significant manner when the team in question is playing in their home stadium (i.e. $\Delta^2 s = 0$).   HA comes from the participation of off-field support players, known as fans, whose function is to increase (or decrease) the auditory and visual noise floors of the match in response to conditions on the field.  The HA effect will bias any joint analysis between home and away data sets. 

These secondary team members participated in a highly limited capacity in the 2020 season. By restricting our analysis to the 2020 season, we hope to mitigate this systematic.  However estimates of the reduction in HA during the pandemic indicate a residual effect unrelated to the off-field support players \citep{price2021effects,krieger2022much, wunderlich2021does}, with estimates as high as 50\% remaining at least for matches of European (i.e. correct) football \citep{mccarrick2021home}.

\begin{figure*}[ht!]
\plotone{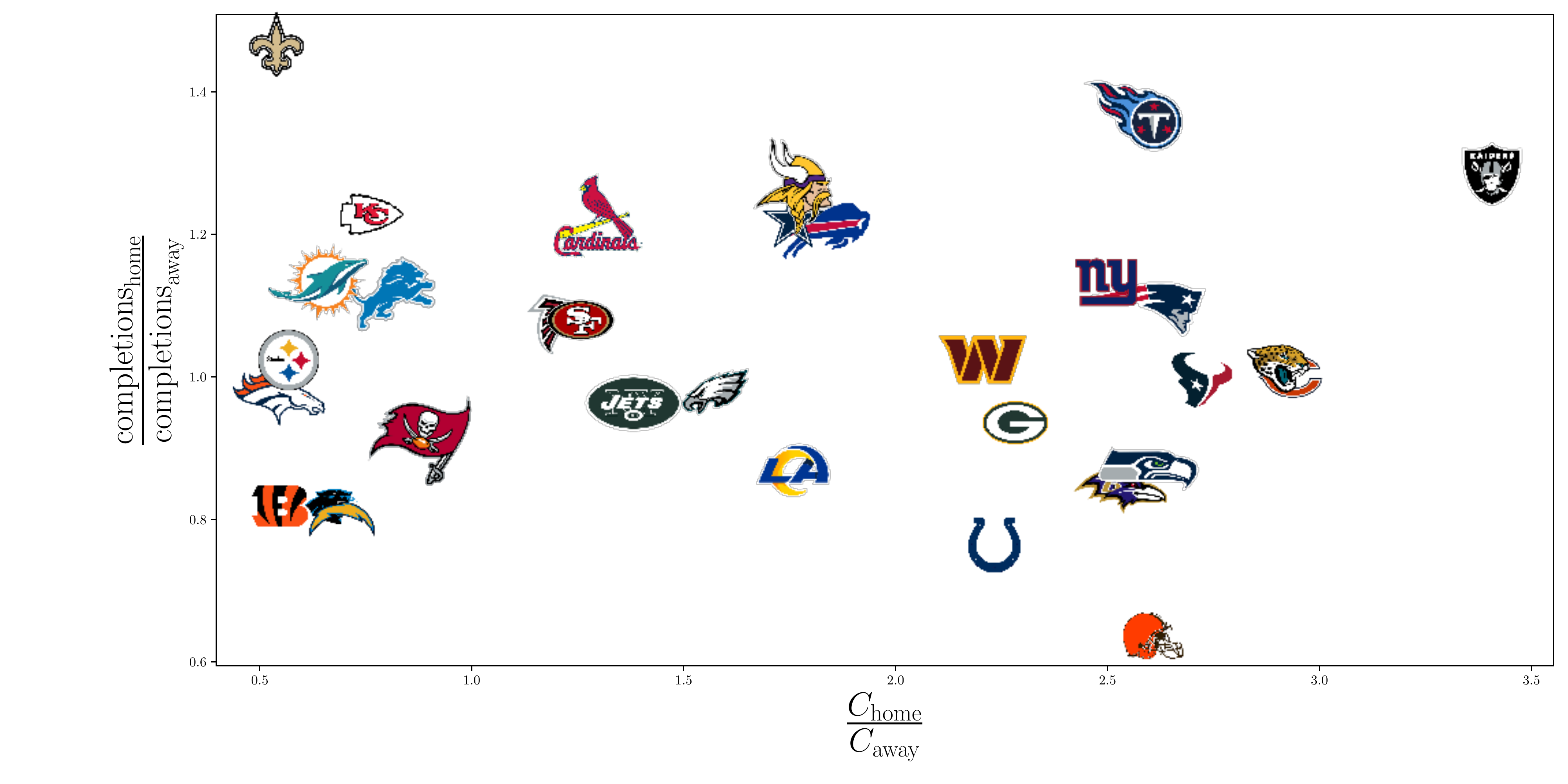}
\caption{Ratio of home to away completions plotted as a function of the home to away color contrasts.  Individual teams are plotted using their logos.
\label{fig:completions}}
\end{figure*}

\subsection{Data}
Each instance of skill realization, or game, can then be collated together to calculate an effective total skill ranking among teams. This is done by playing N games over a season, where the number of games is given by the binomial coefficient, \(\binom{n}{k}\),  defined as,
\[
    \binom{n}{k} = \frac{n!}{k!(n-k)!}
\]
with $n$ corresponding to the number of teams, $N_{\mathrm{team}}=32$, and $k=2$.

Team colors were acquired by a skillful application of googling\footnote{\url{https://teamcolorcodes.com/nfl-team-color-codes/##What_Are_the_RGB_Color_Codes_Used_by_NFL_Teams}}. 

We use the publicly available python package \texttt{nfl-data-py} to generate our dataset. Using this package, we can extract the parameters and results of each play for each game of every NFL season\footnote{Before the modern era of the NFL, the logging of statistics were much less comprehensive and thus not all statistical metrics in \texttt{nfl-data-py} are equally available historically }. To prepare our dataset, we first select a $\rm{T}_{i}$, where $i$ is an integer which ranges from $i = 1 $ to $i = 32$ corresponding to each of the 32 teams in the league. We separate the games of $\rm{T}_{i}$ into home and away categories. For each category we compute the total number of times $\rm{T}_{i}$ was tackled by the opposing team. We exclude plays which ended in a score, turnover or with a player voluntarily sacrificing himself to the defense. We also ignore special team plays, i.e. kickoffs, punts, and field goals. To compute completions we similarly separate the games of $\rm{T}_{i}$ into home and away. We compute the total number of times $\rm{T}_{i}$ completed a forward pass to a member of $\rm{T}_{i}$. We ignore special team plays.

\subsection{Analysis}
We now want to quantify the relationship between the contrast and tactic outcomes.  We do this by taking the ratio of the home and away result of the tactic being considered ($\mathrm{tactic}_{\mathrm{home}} / \mathrm{tactic}_{\mathrm{away}}$) with the ratio of the contrast between the home and away teams ($\mathrm{contrast}_{\mathrm{home}} / \mathrm{contrast}_{\mathrm{away}}$).  If the effect of chromaticity is sufficiently large, than we expect these ratios to be correlated.

\subsubsection{Pearson correlation coefficient}
We restrict our analysis to an investigation of a linear correlation by calculating the Pearson correlation coefficient, hereon simply, the correlation.  Given two data sets, $x$ and $y$, the correlation is given by the ratio of the covariance of the two data sets, divided by the product of their standard deviations,
\begin{align}
    r_{xy} = \frac{\Sigma_{i=1}^N (x_i - \bar{x})(y_i - \bar{y})}{\sqrt{\Sigma_{i=1}^N (x_i - \bar{x}})^2 \sqrt{\Sigma_{i=1}^N (y_i - \bar{y})^2}}
\end{align}
where $\bar{x}$ and $\bar{y}$ are the sample means per each data set, and $\Sigma_{i=1}^N (x_i - \bar{x}^2) = \sigma^2_x$ is the sample variance of the data set, $x$. 

We assume standard errors given by,
\begin{align}
    \sigma^2 = \frac{1-r^2}{N-2}
\end{align}
where $r$ is the calculated correlation coefficient, and N is the number of data points, as above.  We note that this estimate of the uncertainty is only valid under the condition that $x$ and $y$ are random variables. We assume this is sufficiently satisfied that the standard error variance is a good approximation of the true variance.  We assert this without proof and hope no one will notice.

\begin{figure*}[ht!]
\plotone{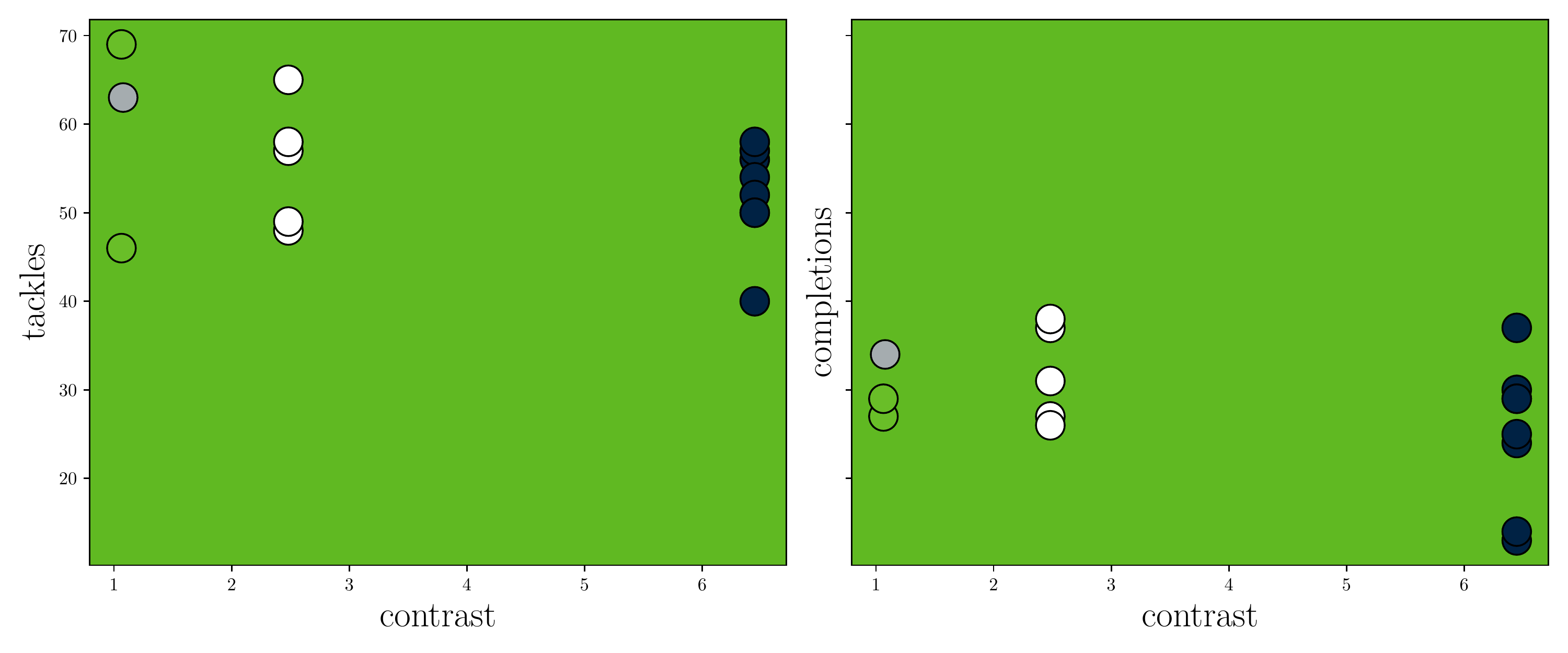}
\caption{The number count of occurrences when a Seahawks player was tackled (left panel) or performed a successful completion (right panel) for each game played in the NFL 2020 regular season, as a function of the uniform color contrast with the field color (background color). Each data point is plotted using the RGB values of the uniform color for the corresponding game.
\label{fig:seahawks}}
\end{figure*}

\section{Results} \label{sec:results}
For the datasets above, we now calculate the Pearson correlation coefficient between the ratios of the tactics and the contrast of home and away games.  These two data sets are shown in Figure\,\ref{fig:tackles} and Figure\,\ref{fig:completions}.  For tackles, we find a correlation value of $r_t=-0.08854763$ .  Similarly, for completions, we find a correlation value of $r_c=-0.02920465$. Since a zero value of the correlation equals no relationship between the variables (with unity indicating a perfect correlation), we can determine that there exists little to no correlation between the two tactics and the color contrast.  The correlation values, and their errors, are tabulated in Table\,\ref{table:pearson}.

\begin{deluxetable}{lcc}
\tablecolumns{3}
\tablewidth{0pt}
\tablecaption{Correlation Coefficients \label{table:pearson}}
\tablehead{
\colhead{Tactics} \vspace{-0.0cm} & \colhead{r} & \colhead{$\sigma_r$} \\}
\startdata
tackles ratio & $-0.0885$ & $0.1819$  \\
completions ratio & $-0.0292$ & $0.1825$  \\
tackles Seahawks & $-0.3606$ & $0.2493$  \\
completions Seahawks & $-0.4049$ & $0.2444$  \\
\enddata
\vspace{-0.0cm}

\end{deluxetable}

\subsection{Investigating Bias for a Single Team}\label{sec:seahawks}
As discussed previously, there does exist non-uniformity in the away colors which is currently unaccounted for in our bulk analysis.  We now turn to a more detailed analysis albeit focusing on a subset of the data. Specifically, we focus on a single team which utilised a wide selection of wardrobe options in the 2020 season: the Seattle Seahawks.

The Seattle Seahawks are the very best American football team in the greater Seattle area, perhaps even the Pacific Northwest, and thus have a reputation to maintain in all manner of things including sartorial choices.  Thus in the 2020 NFL regular season, the Seahawks wore four distinct uniform colors.  In our previous analysis, we neglected this (and other) perturbations away from white.  However, in this additional analysis we now explicitly put the Seahawks colors in by hand.  We then calculate the correlation between the contrast of \textit{each game} with the number of successful tactics as outlined in Section\,\ref{sec:methodology}. The data for this secondary analysis is shown in Figure\,\ref{fig:seahawks}.

The analysis yields values of $r_t=-0.3606$ and $r_c=-0.4049$ for tackles and completions, respectively.  Thus we conclude there is moderate evidence for a negative correlation between the color contrast of the uniform and the number of times that team is tackled or had a successful completion per game.  However, we note that the errors bar are large, leading to a signal-to-noise ratio (SNR) of nearly unity.  We caution that further analysis is needed which we intend to address in exactly one year from now. Maybe.

\section{Conclusion} \label{sec:conclusion}
In this work we investigated the possible effect of chromatic variation of uniform colors among NFL American football teams on the outcomes of scored games. Two datasets are considered in this analysis. The tackles (completions) dataset is composed of the ratio of the number of home tackles (completions) to away tackles (completions) over the 2020 NFL regular season. This ratio is then combined with the ratio of the color contrasts of each team's uniform against the field color for both the home uniform color and the away uniform color. We then investigated the relationship between these ratios by calculating the Pearson correlation coefficient.  We find little correlation between tactics performance and uniform color.

We then perform an additional analysis that sidesteps the systematic bias due to the variations of uniform color per game not fully modeled in our primary analysis.  This is done by conducting a finer-grained analyis on a particular team, the Seattle Seahawks, which played wearing four unique colors during the 2020 seasons considered here.  We find moderate evidence for a negative correlation between how often a Seahawks player was tackled by the opposing team and the color contrast of the uniform color for that game, and similarly we find a negative correlation for completions as well.  However, we caution that the large error bars associated with this correlation due to the limited number of samples (only $N=18$ games are played per team in a typical season), further work is needed. Of note is that the correlations are negative for both tactics.  This doubly negative correlation could hint at the necessity of improved models for how contrast affects spheroidal ball matches.

Given the large errors on the secondary analysis, we conclude that there is still little evidence of an effect of team uniform colors on the outcomes of two specific football tactics.  However, future studies could find an effect on other tactics, or perhaps a holistic analysis per game would indicate an effect.  Additionally, since the Pearson correlation coefficient only quantifies the linear correlation, it is still possible that color contrast has a higher-order effect.  However, it is likely to be highly subdominant.  In summary, it appears that chromaticity effects cannot explain why the Browns are not very good.
\\
\\
\\
\\
All code written for this analysis and data sets are available for public use and can be accessed at \url{https://github.com/lisaleemcb/chromaball}.

\section*{Acknowledgements}
The authors would like to thank Reed Fife for several stimulating discussions on this topic, and also to Kelly Foran for raising valuable key points that helped sharpen the scope of our analysis. Additionally, the authors would like to thank McGill University, and Canada, for helping to invent American football. Finally, the authors would like to humbly acknowledge the work of the NFL commissioner, Roger Goodell, and the whole NFL organization, for asserting that $\mathrm{football} \gg \mathrm{pandemic}$. Without his efforts to play football during the shutdown, this dataset and thus entire work, would not have been possible.  We love you, please don’t sue us.  Go Patriots, and go Hawks!

\vspace{5mm}


\software{numpy \citep{harris2020array},  
          nfldata \citep{cooperdff}}

\bibliography{chromaball}{}
\bibliographystyle{aasjournal}



\end{document}